\documentclass[prb,aps,twocolumn,superscriptaddress]{revtex4}
\usepackage{graphicx}
\usepackage{natbib}
\usepackage{mathrsfs}
\usepackage{amsmath}
\usepackage{ulem}
\usepackage{graphicx}

\begin{document}

\title{Robust short-range-ordered nematicity in FeSe evidenced by high-pressure NMR }

\author{P. S. Wang}
\affiliation{Department of Physics and Beijing Key Laboratory of
Opto-electronic Functional Materials $\&$ Micro-nano Devices, Renmin
University of China, Beijing, 100872, China}

\author{P. Zhou}
\affiliation{Department of Physics and Beijing Key Laboratory of
Opto-electronic Functional Materials $\&$ Micro-nano Devices, Renmin
University of China, Beijing, 100872, China}

\author{S. S. Sun}
\affiliation{Department of Physics and Beijing Key Laboratory of
Opto-electronic Functional Materials $\&$ Micro-nano Devices, Renmin
University of China, Beijing, 100872, China}

\author{Y. Cui}
\affiliation{Department of Physics and Beijing Key Laboratory of
Opto-electronic Functional Materials $\&$ Micro-nano Devices, Renmin
University of China, Beijing, 100872, China}

\author{T. R. Li}
\affiliation{Department of Physics and Beijing Key Laboratory of
Opto-electronic Functional Materials $\&$ Micro-nano Devices, Renmin
University of China, Beijing, 100872, China}

\author{Hechang Lei}
\affiliation{Department of Physics and Beijing Key Laboratory of
Opto-electronic Functional Materials $\&$ Micro-nano Devices, Renmin
University of China, Beijing, 100872, China}

\author{Ziqiang Wang}
\affiliation{Department of Physics, Boston College,
Chestnut Hill, Massachusetts 02467, USA}

\author{Weiqiang Yu}
\email{wqyu_phy@ruc.edu.cn}
\affiliation{Department of Physics and Beijing Key Laboratory of
Opto-electronic Functional Materials $\&$ Micro-nano Devices, Renmin
University of China, Beijing, 100872, China}
\affiliation{Department of Physics and Astronomy, Shanghai Jiaotong
University, Shanghai 200240, China and \\ Collaborative Innovation Center
of Advanced Microstructures, Nanjing 210093, China}

\begin{abstract}

We report high-pressure $^{77}$Se NMR studies on FeSe single crystals that reveal a prominent inhomogeneous NMR linewidth broadening
upon cooling, with the magnetic field applied along the tetragonal
[110] direction. The data indicate the existence of
short-range-ordered, inhomogeneous electronic nematicity,
which has surprisingly long time scales over milliseconds.
The short-range order survives temperatures
up to $8$ times the structural transition temperature,
and remains robust against pressure, in contrast
to the strong pressure-dependence of the orbital ordering, structural transition, and the ground state magnetism.
Such an extended region of static nematicity in the
($P$,$T$) space of FeSe indicates an enormously large fluctuating
regime, and provide fresh insights and constraints to the understanding
of electronic nematicity in iron-based superconductors.

\end{abstract}

\maketitle

The $C_4$ rotational symmetry breaking electronic nematicity~\cite{Emery_Nematic_Nature_1998} has been widely observed in the cuprates~\cite{Lavrov_PRL_2002,Keimer_Science_2008,Vojta_AdvP_2009,Taillefer_Nature_2010}
and the iron-based superconductors
(FeSCs)~\cite{Fisher_Science_2010, Davis_Science_2010,FengDL_PRB_2012}.
In the FeSCs, the onset of the long-range nematic order is usually
companied by a tetragonal-to-orthorhombic structural transition.
Since the large anisotropy of the in-plane resistivity
in the nematic phase cannot be accounted for by the rather small
anisotropy of the in-plane lattice parameter, this structural transition is likely triggered by the nematic order of the electronic state~\cite{Fisher_Science_2010,Prozorov_PRB_2010}. However, the physical origin of the electronic nematicity
is still highly debated in the FeSCs~\cite{Fenandes2014}.
In the iron pnictides, the stripe-ordered magnetism~\cite{DaiPC_Nat_2008} sets in at or just below the nematic ordering transition, fueling the debate between a magnetic-driven spin-nematic~\cite{Kivelson_PRB_2008,Xu_nematic_PhysicaC_2012} and the ferro-orbital order ~\cite{Phillips_PRB_2009,KuW_PRL_2009,Buchner_NatM_2015,Bohmer_PRL_2015}
scenarios.

Recent studies on bulk FeSe~\cite{WuMK_PNAS_2008} find that the nematic order~\cite{Takahashi_PRL_2014,Prozorov_PRL_2016} occurs
simultaneously with orbital ordering~\cite{Takahashi_PRL_2014,Coldea_PRB_2015,DingH_PRB_2015}
at the structural transition $T_S\sim$~90~K~\cite{McQueen_PRL_2009}, whereas the stripe-ordered magnetism is absent at the ambient pressure~\cite{Takano_PhysC_2010}. These findings support the orbital-order driven electronic nematicity in bulk FeSe, with possible momentum space anisotropy \cite{DingH_PRB_2015,WangZQ_PRB_2016}.
However, there are also controversies on the origin of nematicity in FeSe. It was proposed that the absence of the striped magnetic order in FeSe could be caused by the competing tendencies of magnetism~\cite{DHLee_NatP_2015,QSi_PRL_2015,Valenti_NatP_2015,
TXiang_PRB_2016}. Under pressure, the temperature of the structural transition and the nematic order is first suppressed at low pressures,
but rises again with an emergent antiferromagnetic order~\cite{Khasanov_PRL_2010,Cheng_NatC_2016,Bohmer_NatC_2016}
that is confirmed to be the stripe type~\cite{YuWQ_PRL_2016}.
%These may be consequences of strong couplings between nematicity and other orders that also breaks the $C_4$ symmetry, while the origin of nematicity again becomes an open question.
One way to test the different scenarios is
to study nematic responses beyond the parameter regimes of other types of intervening ordering~\cite{Xu_nematic_PhysicaC_2012,Fradkin_prb2014}.
Indeed, in iron pnictides, anisotropic resistivity studies revealed a
Curie-Weiss type singularity above $T_S$ when an
external uniaxial stress is applied~\cite{Fisher_Science_2010, Schmalian_NematicFluc_PRL_2010};
inelastic neutron scattering~\cite{DaiPC_Neutron_PRB_2011} and
STM measurements~\cite{Pasupathy_NatP_NaFeAs_2014} found signatures of nematicity at finite energies; and nuclear magnetic resonance (NMR)
studies observed features of line splitting and inhomogeneous spin-lattice relaxation rates
~\cite{ZhengGQ_BaFeNiAs_NatC_2013,Matsuda_BaFeAsP_JPSJ_2015,Curro_PRL_2016}.
Recent ARPES~\cite{DingH_PRB_2015}, optical-pumped conductivity
measurements~\cite{Vasiliev_Pump_arXiv_2016} and Raman scattering~\cite{Massata_PNAS}
in FeSe also revealed nematicity signatures well above the structural transition or orbital ordering temperature. However, the time/energy scale of
these nematic responses at high-temperatures remains elusive.

In this paper, we report a high-pressure NMR study on FeSe,
whose narrow $^{77}$Se NMR line turns out to be essential
for resolving the nematic response at high-temperatures. We found a
prominent increase of the linewidth of the $^{77}$Se spectra upon cooling toward $T_S$,
with field applied along the tetragonal [110] direction,
but not along the [100] direction.
This indicates an in-plane anisotropy of the Knight shift in the system,
since $^{77}$Se is a spin-1/2 nucleus which only detects magnetic responses.
By comparing with $1/^{77}T_2$, a static, spatially inhomogeneous distribution of nematic response is concluded with
time scales over milliseconds, consistent with short-range-ordered (SRO) nematicity.
The static nematicity survives temperatures up to
$8\times T_S$ and does not change with pressure up to 2.4 GPa.
Our observation of robust SRO nematicity against temperature 
and pressure, in contrast to the prominent pressure-dependence of
orbital order and magnetic order, provides new insights and strong constrains on the theory of electronic nematicity in FeSe.

The FeSe single crystals were synthesized by an
assisted-flux method, whose high quality
was demonstrated by our previous high-pressure
NMR study on its magnetic structure~\cite{YuWQ_PRL_2016}.
The sample was loaded in a piston cell
with Daphne oil 7373 as the pressure medium, and
the cell was heated to~$\sim$~80~$^{\circ}$C when pressurizing
to above 2 GPa for better pressure hydrostaticity~\cite{YuWQ_PRL_2016}.
The low-temperature pressure was determined at 5 K by the
$^{63}$Cu NQR frequency of Cu$_2$O powders loaded in the pressure
cell~\cite{Thompson_Cu2O_NQR_HP}.
We verified that pressure barely changes with temperature below 100 K.
For NMR measurements, a constant field of 10.3~T
was primarily applied along one tetragonal [110] direction of the sample,
which becomes the $a$ or the $b$-axis in
the twinned orthorhombic phase~\cite{Ma_NaFeAs,Imai_PRL_2012,Buchner_NatM_2015,Bohmer_PRL_2015}.
The $^{77}$Se NMR signals were accumulated with the standard
spin-echo technique. The spin-spin relaxation rates $1/^{77}T_2$
are measured by the standard Hahn spin-echo sequence,
and the decay is nicely fit by a single exponential function
of the interpulse delay time.

\begin{figure}[t]
\includegraphics[width=8.5cm, height=8cm]{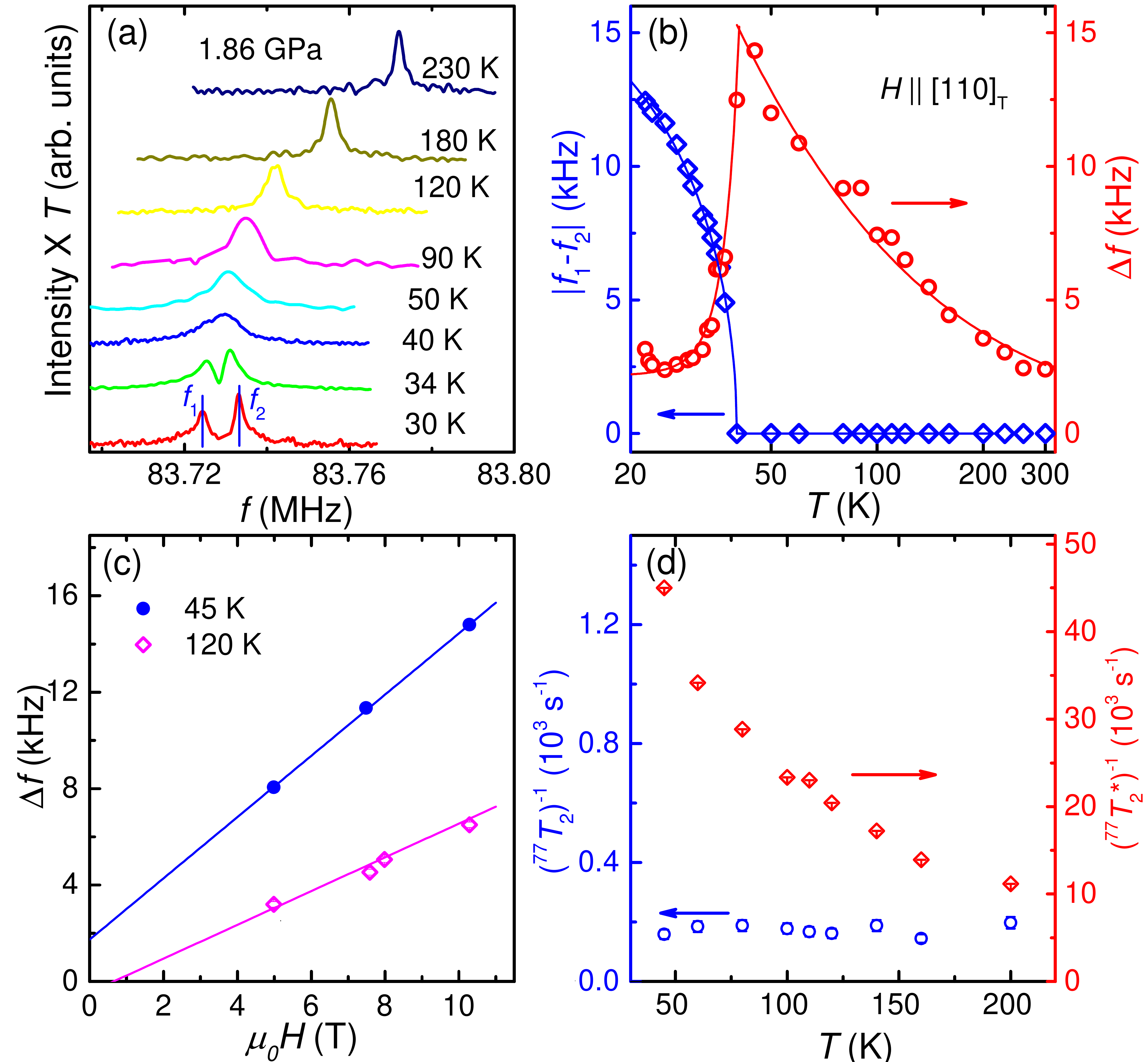}
\caption{\label{spec1}. $^{77}$Se NMR data measured under 1.86 GPa pressure.
(a) The $^{77}$Se NMR spectra at selected temperatures.
(b) The line separation ($|f_1-f_2|$) of two NMR lines
and the FWHM (${\Delta}f$) as functions of temperatures,
with $x$-axis in a log-scale.
Below $T_S$, only the data of the $f_2$ line are shown.
The solid lines are eyeball guides to the data points.
(c) The ${\Delta}f$ as functions of fields.
(d) The $1/^{77}T_2$ and the $1/^{77}T_2^{*}$ (see text)
as functions of temperatures.
}
\end{figure}

The $^{77}$Se NMR spectra at a typical pressure of 1.86 GPa
are shown in Fig.~\ref{spec1}(a). At 230 K, a single-peaked NMR
line is observed. Upon cooling,
the spectra first shift to low frequencies with an obvious increase of the NMR linewidth. Further cooling below 40 K, the spectra split into
double peaks (with peak frequency $f_1$ and $f_2$ respectively),
which is an evidence for the tetragonal-to-orthorhombic
structural transition, with twinned orthorgonal structure
domains below $T_S$~\cite{Ma_NaFeAs}.
The frequency splits between the two peaks, denoted by $|f_1-f_2|$, is proportional
to the field~\cite{Buchner_NatM_2015,Bohmer_PRL_2015,YuWQ_PRL_2016}, illustrating
a two-fold anisotropy of the Knight shift in the orthorhombic
phase. Fig.~\ref{spec1}(b) plots $|f_1-f_2|$
as a function of temperature, which behaves as an order parameter of
nematicity\cite{Buchner_NatM_2015,Bohmer_PRL_2015} below
$T_S\approx$~40~K. This structural transition corresponds to
the orbital order~\cite{Takahashi_PRL_2014,Coldea_PRB_2015,DingH_PRB_2015}
and the nematic order~\cite{Prozorov_PRL_2016} revealed at the ambient pressure.

\begin{figure}
\includegraphics[width=8.5cm, height=5.5cm]{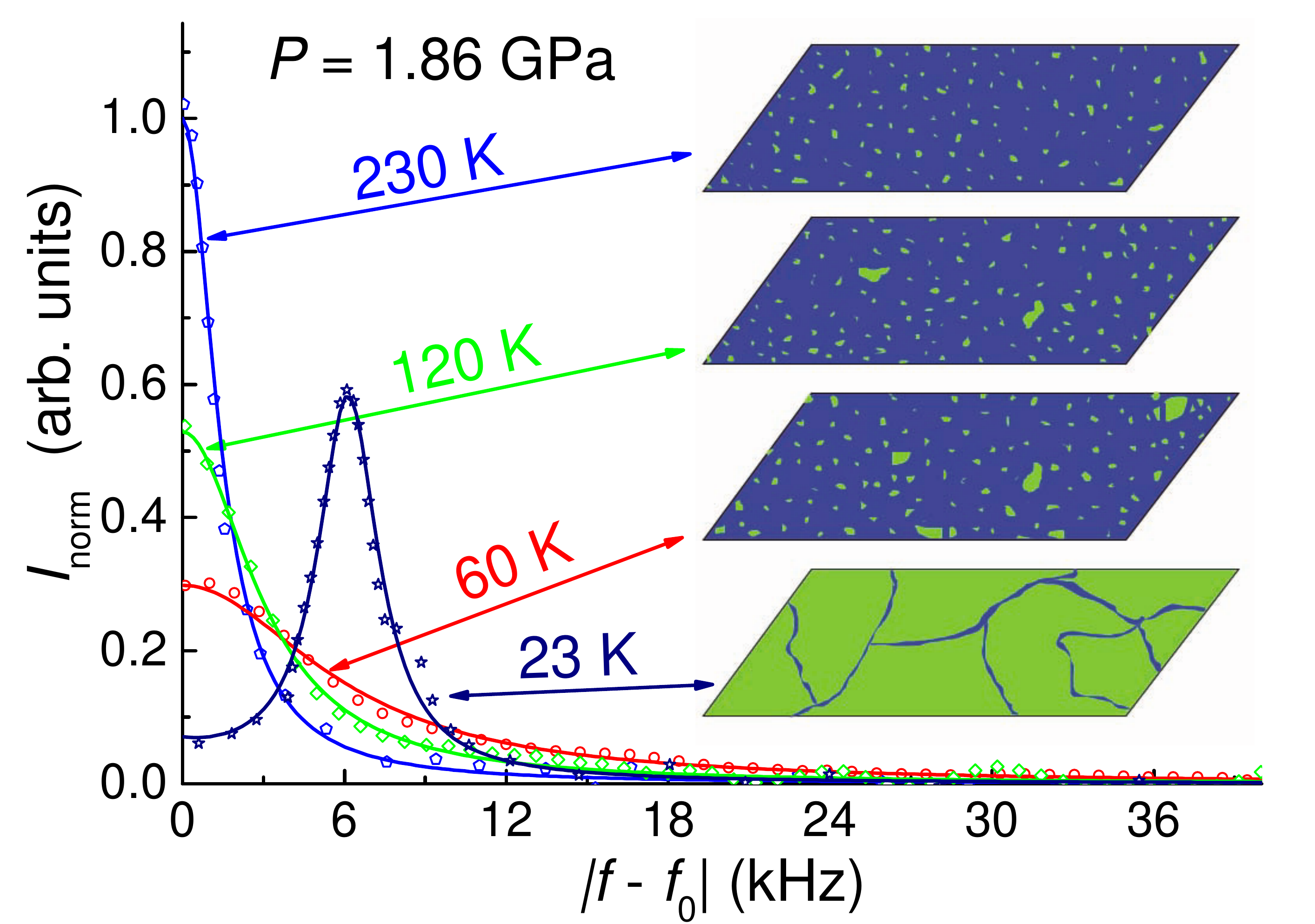}
    \caption{\label{nematic}
The relative spectra intensity deduced from the
NMR lineshapes (see text) at four typical temperatures
with $P=$ 1.86 GPa. Here the contributions from
the negative and the positive $f-f_0$ are averaged.
The solid lines are Lorentz function
fit to the data. The color maps are schematic drawings
of sample regions with (green color) and without
(blue color) static nematicity.
}
\end{figure}

The full-width-at-half-maximum (FWHM) of each peak, denoted by ${\Delta}f$, is obtained by the Lorentz fit to the line and plotted
as a function of temperature in Fig.~\ref{spec1}(b). It first increases
from 2.5~kHz at $T=$~300~K to 14~kHz at 45~K, and then drops sharply
below $T_S$, reaching a small value of~$\sim$2.5~kHz again at 25~K.
Detailed NMR lineshape analysis for temperatures from 230 K down to $T_S$ reveals that i) each line above $T_S$ has a single peak
(Fig.~\ref{spec1}(a)) and is well fit with
a simple Lorentz function (see Fig.~\ref{nematic}).
By contrast, the double-line fitting with six parameters, used in iron
pnictides~\cite{ZhengGQ_BaFeNiAs_NatC_2013,Matsuda_BaFeAsP_JPSJ_2015},
does not give converging fitting parameters in our FeSe system.
ii) The FWHM ${\Delta}f$ at $T\approx$~45~K is comparable
to the line split $|f_1-f_2|$ at 25~K (13~kHz);
iii) The ${\Delta}f$ at $T=$~200~K is comparable to the FWHM of each split peak at 25~K, but increases by about four times at 45~K.

\begin{figure}
\includegraphics[width=8.5cm, height=5.5cm]{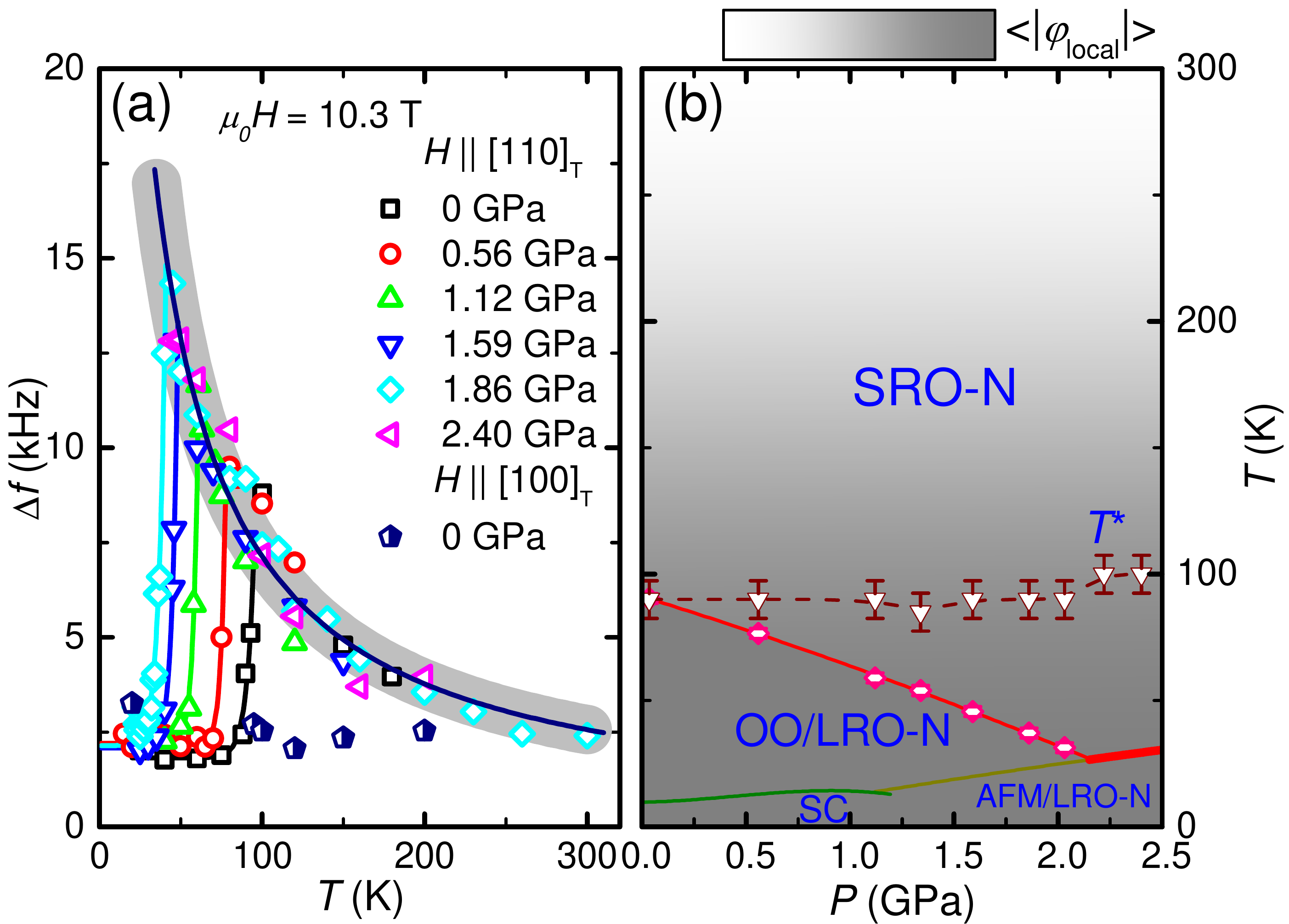}
\caption{\label{pd}
(a) The FWHM as functions of temperature at selected pressures.
The navy solid line presents one Curie-Weiss function fit
to all data sets above $T_S$, with the gray color covering all
data points. The FWHM with field applied along the
tetragonal [100] direction is also plotted under zero pressure.
(b) The ($P$, $T$) phase diagram of FeSe. The SRO nematic phase (SRO-N)
 is determined from this study. The LRO nematic phase (LRO-N), the
 orbital-ordered phase (OO), the stripe-antiferromagnetic phase (AFM),
 the superconducting phase (SC), and the onset temperature
 of low-energy spin fluctuations ($T^*$) are adapted from Ref.~\citenum{YuWQ_PRL_2016}.
The grey color maps the averaged value of $|{\varphi}_{local}|$ (see text).
}
\end{figure}

The ${\Delta}f$ is further measured as a function of field
at typical temperatures above $T_S$ as shown in Fig.~\ref{spec1}(c).
A nearly linear-field dependence is clearly
seen, which indicates that the line broadening is caused
by a distribution of the local susceptibility
rather than a field-induced effect,
similar to the linear field-dependence of the line splitting
below $T_S$.
Given the narrow linewidth below $T_S$, and the close values between
the $|f_1-f_2|$ (below $T_S$) and ${\Delta}f$ (just above $T_S$),
the high-temperature line broadening
should be considered as a mixing of nematicity with
distributed amplitudes across the sample.
In fact, since the line broadening only
occurs with field applied along the tetragonal [110] direction,
and not along the [100] direction (see Fig.~\ref{pd}(a)),
such an in-plane anisotropy is a direct evidence for nematicity.

To check whether the nematicity above $T_S$
is a dynamic effect or a static phenomenon,
the spin-spin relaxation rate $1/^{77}T_2$ is measured
and plotted as a function of temperature
in Fig.~\ref{spec1}(d). $1/^{77}T_2$ stays nearly constant
in the measured temperature range.
By contrast, the $1/^{77}T_2^*$ ($={\pi}$${\Delta}f$) increases by 25 to 250 times of $1/^{77}T_2$
when cooled with temperatures from 200 K down to 45 K (Fig.~\ref{spec1}(d)).
In general, $1/T_2^*=1/T_2+\pi\gamma{\Delta}H$ for
s=1/2 nuclei, where $1/T_2$ detects the homogenous broadening of spectra due
to dynamical magnetic fluctuations, and ${\Delta}H$ is the spatial field
inhomogeneity which has a time scale longer than $T_2$ at the nuclear sites.

Such a strongly temperature-dependent $1/^{77}T_2^*$
concomitant with the small and constant $1/^{77}T_2$
indicates a spatially inhomogeneous electronic environment
across the sample, rather than a dynamical fluctuating one,
responsible for the line broadening.
The lower bound of the time scale for this spatial
inhomogeneity is thus given by the value of
$^{77}T_2$, that is, 5~milliseconds for all
temperatures (see Fig.~\ref{spec1}(d)), since
no change of the spin dynamics within this time scale
is seen upon cooling. With such a long time scale in an
electronic system, the nematicity response should be
taken as a static phenomenon for temperatures over 200~K.

In the following, we apply a spectra analysis to reveal a
possible real-space distribution of the SRO nematicity.
Since the line splitting below $T_S$ is proportional to
the amplitude of the nematic order parameter~\cite{Buchner_NatM_2015,Bohmer_PRL_2015}, we introduce
a {\it local} ``order'' parameter of nematicity, $\varphi_{local}$,
measured by the distance from the NMR resonance frequency ($f$) to the
center of the spectrum ($f_0$) at each temperature, i.e. $\varphi_{local}=f-f_0$.
The $^{77}$Se spectrum thus maps out the relative volume fraction
at each $\varphi_{local}$ which varies across the sample.
Following this, the relative spectral intensity $I_{norm}$ as functions
of $|f-f_0|$ is presented in Fig.~\ref{nematic} for typical temperatures, with the total spectral weight normalized to the same value for all temperatures.
From 230 K down to 23 K, each data set can be fit with a
simple Lorentzian (solid lines). Above $T_S$, the spectral
weight remains peaked at $|f-f_0|=0$, but broadens toward higher
$|f-f_0|$ with a wider distribution upon cooling.
Below $T_S$, a narrow peak is formed at a finite $|f-f_0|$.
We emphasize that our determination of the inhomogeneous
SRO nematicity takes advantage of the narrow NMR linewidth
far below $T_S$.

The above spectral distribution corresponds to the
volume distribution of $\vert\varphi_{local}\vert$ at each temperature.
The wide distribution of local ``order" at temperatures
up to 200 K should already indicate the formation of inhomogeneous
nematicity, or bubbles of the nematic phase. It is reasonable to assume that the
amplitude of $\varphi_{local}$ increases with the domain size of the nematic phase,
when the long-range-ordered (LRO) nematicity is not formed.
Therefore, the wide distribution of $\varphi_{local}$ above $T_S$
indicates a form of SRO nematicity with non-uniform domain sizes across the sample.
On the other hand, the formation of LRO
nematicity is shown by the nearly divergent nematic susceptibility,
which coincides with the structural transition and the orbital order
at $T_S$, as shown at ambient pressure~\cite{Coldea_PRB_2015,Prozorov_PRL_2016}.
With this, schematic drawings of the nematic phases at different temperatures are
presented as color maps in Fig.~\ref{nematic}: inhomogeneous,
local static nematic order already occurs far
above $T_S$ (shown at 230 K), with decreasing volume fraction for large domains;
upon cooling, the volume fraction of the large domains grows
until a uniform, LRO phase develops sharply below $T_S$ (shown at 23 K).

\begin{figure}
    \includegraphics[width=8.5cm, height=8cm]{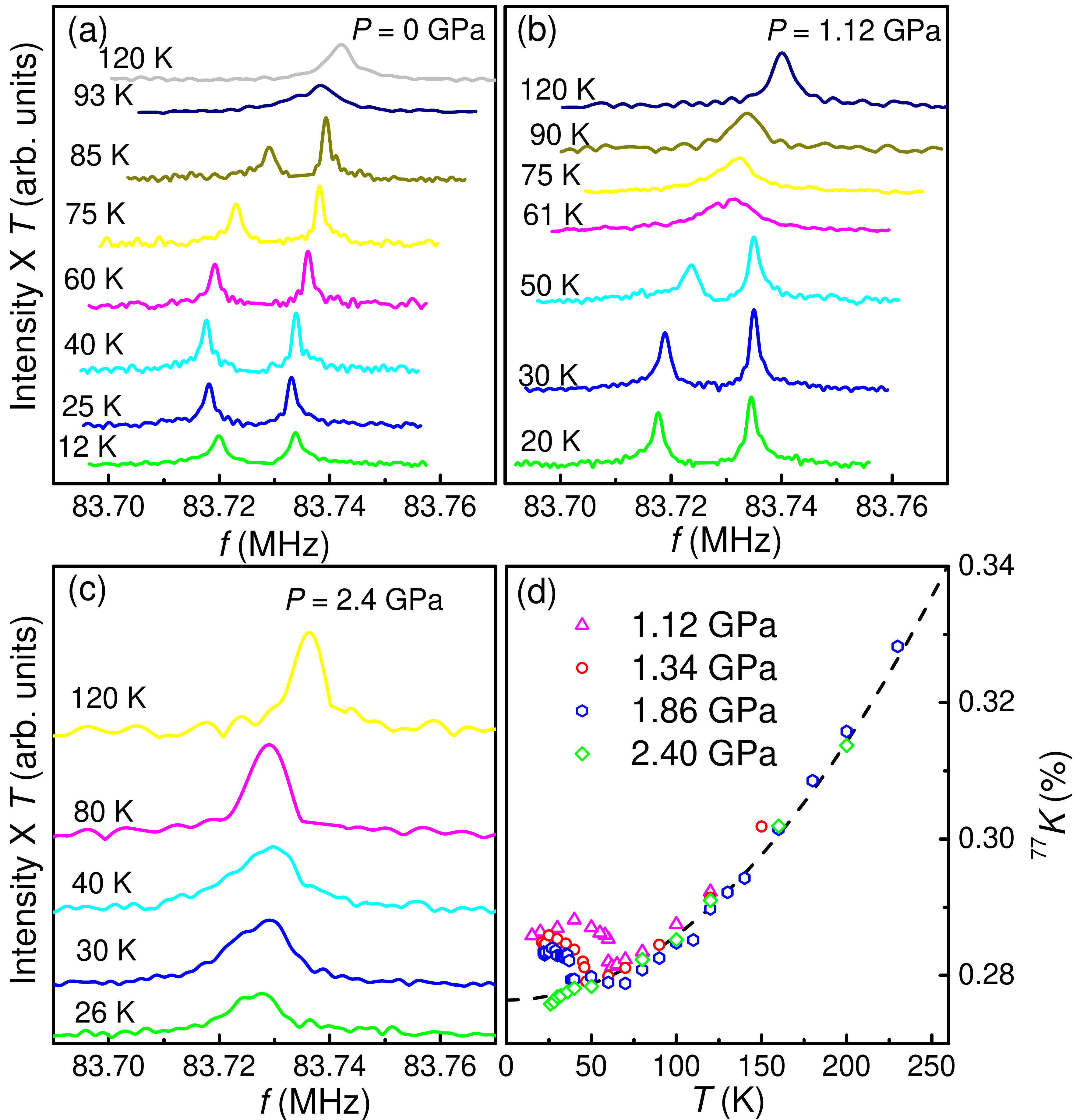}
    \caption{\label{specp} (a)-(c): The $^{77}$Se NMR spectra at typical temperatures,
measured under the pressure of (a) $P=$0 GPa, (b) 1.12 GPa,  and (c) 2.4 GPa.
(d) The Knight shift $^{77}K$ determined from the peak frequency of the NMR lines.
Below $T_S$, only the data of the high-frequency line are plotted.
}
\end{figure}

The NMR spectra at three other pressures are further shown in Figs.~\ref{specp}(a)-(c).
The structural transitions are observed at 0 GPa and 1.12 GPa as well,
by the line splitting at low temperatures.
The NMR Knight shift $^{77}K$, defined as $^{77}K=(f-^{77}{\gamma}B)/^{77}\gamma B$,
with the gyromagnetic ratio $^{77}{\gamma}\sim$~8.118~MHz/T and under
the external field $B$, are presented in Fig.~\ref{specp}(d).
The $T_S$ can also be identified by the kink features in the $^{77}K$.

The FWHM of the NMR spectra at different pressures are summarized
as functions of temperature in Fig.~\ref{pd}(a).
For all pressures, large increases of the linewidth are observed
when the sample is cooled toward $T_S$.  Below $T_S$, a
similar ${\Delta}f$ $\sim$~2.5~kHz is achieved at the lowest temperatures.
In fact, all data follow an identical temperature dependence
above $T_S$, which can be fit
by a single Curie-Weiss function, ${\Delta}f=a/(T-\theta)+b$,
with $\theta=-12\pm2$K and $b$ negligibly small,
as shown by the solid navy line in Fig.~\ref{pd}.
This remarkable result indicates that the amplitude of SRO nematicity
at a given temperature is nearly independent of pressure.

As an important check, we measured the NMR spectra (not shown) with field
applied along the tetragonal [100] direction at ambient pressure.
The obtained linewidth is shown in Fig.~\ref{pd}(a) which remains a
small constant with ${\Delta}f\sim$ 2.5~kHz down to $T_S$. The absence of
line broadening upon cooling under this field orientation
(the orthorhombic [110]) direction) verifies the same nematic
orientation as below $T_S$~\cite{Buchner_NatM_2015}.

For comparison, the obtained evidence of SRO nematicity at
the ambient pressure is consistent with the
ARPES~\cite{DingH_PRB_2015} and optical-pump conductivity~\cite{Vasiliev_Pump_arXiv_2016}
data, and further reveals a very low energy-scale of nematicity by our
observation of its long time scale.
The nematic susceptibility at ambient pressure was reported to follow
a Curie-Weiss (CW) form~\cite{Fisher_NematicSus_Science_2012},
but with a positive $\theta$ (${\approx}T_S$)~\cite{Coldea_PRB_2015,Prozorov_PRL_2016}.
The difference in the $\theta$ values can be understood by the
fact that our ${\Delta}f$ measures the spatial distribution of the local order parameter, whereas the nematic susceptibility
measures the temporal correlations associated with the low-energy dynamics.

Finally, we sketch a phase diagram with the
SRO nematic phase determined from the current study in Fig.~\ref{pd}(b).
A schematic color map is also plotted in Fig.~\ref{pd}(b) to illustrate
the averaged value of $|{\varphi}_{local}|$ (or $|f-f_0|$)
by the spectral weight (see Fig.~\ref{nematic}).
The orbital ordered phase, the magnetically ordered phase,
and the superconducting phase from previous NMR experiments~\cite{YuWQ_PRL_2016} are also presented
in the studied pressure regime. The $<|{\varphi}_{local}|>$ grows for all pressures
upon cooling below 200~K, which suggests that the onset temperature of the SRO nematicity is enormously high.
In particular for $P=$ 2.15 GPa, the linewidth broadening is
also seen at 200 K, even though the structural transition and
the stripe-order magnetism emerge only below 25 K~\cite{YuWQ_PRL_2016}.
This marks an unusual high onset temperature ($8\times T_S$) of SRO nematicity.

The observation of the SRO nematicity with very long
time scales at such high temperatures is unexpected,
since it requires slow short-range fluctuations.
It is generally known that, close to the LRO phase, the short-range fluctuations can be slowed down by pinning effects
from quenched disorder and/or residual
stress after the crystal growth~\cite{LiY_PRL_2015}.
However, the very small ${\Delta}f$ observed at 200~K suggests
that the quenched disorder/stress should be
very weak in FeSe. In order to account for the
static local nematicity at such high temperatures, a
strong pinning effect to nematicity may
have to be introduced, even in the presence of weak disorder/stress.

We should point out that our spin recoveries from
the spin-lattice relaxation ($1/^{77}T_1$)
are well fit by a single exponential function
across the entire spectra above $T_S$~\cite{YuWQ_PRL_2016}.
It is known that the in-plane spin dynamics is
anisotropic in the stripe phase~\cite{Takigawa_JPSJ_2008,Hosono_PRB_2010}.
The observation of a single component of $^{77}T_1$ above $T_S$
could be a challenge to the scenario of static nematicity.
However, even at the ambient pressure,
the anisotropy of $1/^{77}T_1$ is only detectable at
temperatures far below $T_S$\cite{Buchner_NatM_2015}, which suggests that
the anisotropy of $1/^{77}T_1$ is not a sensitive probe of nematicity
in FeSe.

That the observed SRO nematicity or nematic fluctuation
is not affected by pressure in bulk FeSe
is striking and further constrains microscopic theories
on the nature of electron nematicity in FeSCs.
Recently, it was proposed that the nematicity in FeSe
may be caused by local Hund's rule couplings~\cite{KuW_PRL_2015},
or by interatomic Coulomb repulsion~\cite{WangZQ_PRB_2016}.
It remains to be seen how such microscopic interaction
parameters, as well as the wavefunction overlap that
governs the electronic structure, are affected by hydrostatic pressure.
It is worthwhile to note that the low-energy spin fluctuations in FeSe
become prominent below a specific temperature $T^*$
at each pressure~\cite{Buchner_NatM_2015,YuWQ_PRL_2016}.
Interestingly, $T^*$ also does not change much with pressure,
as shown in Fig.~\ref{pd}(b).
These similar pressure behaviors draw a possible correlation between
nematicity and low-energy spin fluctuations, and challenge the
scenario of the orbital-driven nematicity~\cite{Buchner_NatM_2015}.

By contrast, the onset temperatures of the LRO nematicity, the orbital ordering,
and the magnetic orderings change dramatically with pressure (Fig.~\ref{pd}(b)).
In particular, the LRO nematicity occurs just below the
orbital ordering temperature at low pressures, and below the
magnetic transition temperature at high pressures~\cite{YuWQ_PRL_2016,Bohmer_PRL_2015},
We think that these coincidences may be understood on
a mean-field level, where the formation of the LRO nematicity is
helped by other types of orderings which also break the $C_4$ symmetry.

In summary, we observed a prominent NMR
linewidth broadening upon cooling in bulk FeSe
in a large pressure range, which is a
direct evidence for the existence of
inhomogeneous, SRO nematicity.
The time scale for the SRO nematicity is surprisingly
long, over milliseconds even at temperatures
far above the LRO phase.
The SRO nematicity also stays robust against pressure,
despite of the dramatic change of the ground-state properties.
Our results also draw a possible correlation between nematicity
and the low-energy spin fluctuations whose onset temperature
also barely changes with pressure.
These distinctive electronic properties will help understand
the microscopic origin of nematicity and its relation to
magnetism in the iron-based superconductors.

We acknowledge encouraging discussions with Prof. Rong Yu
and Wei Ku. Work at Renmin University of China is supported
by the National Natural Science Foundation
of China (NSFC) (Grant Nos. 11374364 and 11574394), the Ministry of
Science and Technology of China (Grant Nos. 2016YFA0300504), and
the Fundamental Research Funds for the Central Universities and
the Research Funds of Renmin University of China (Grant Nos. 15XNLF06 and 15XNLQ07).
ZW is supported by U.S. Department of Energy, Basic Energy Sciences Grant DE-FG02-99ER45747.

\end{document}